\documentstyle[a4,12pt,epsf]{article}
\def\be{\begin{equation}}
\def\ee{\end{equation}}
\def\bq{\begin{eqnarray}}
\def\eq{\end{eqnarray}}
\begin{document}
\thispagestyle{empty}
\setcounter{page}{0}
\setcounter{page}{0}
\begin{flushright}
WUE-ITP-97-001\\
MPI-PhT/97-5\\
Zurich Uni-31/96\\
Technion-ph-96-24\\
January, 1997
\end{flushright}
\vspace*{\fill}
\begin{center}
{\Large\bf QCD estimate of the long-distance effect in 
$B \rightarrow K^* \gamma$ }\\
\vspace{2em}
\large
A. Khodjamirian$^{a,1}$, R. R\"uckl$^{a,b}$\\
\vspace{1em}
{$^a$ \small Institut f\"ur Theoretische Physik, Universit\"at 
W\"urzburg, D-97074 W\"urzburg, Germany }\\
{$^b$ \small Max-Planck-Institut f\"ur Physik, Werner-Heisenberg-Institut, 
D-80805 M\"unchen, Germany}\\
\vspace{2em}
G. Stoll$^c$, D. Wyler$^{c,d}$\\
\vspace{1em}
{$^c$ \small Institut f\"ur Theoretische Physik, Universit\"at 
Z\"urich, CH-8057 Z\"urich, Switzerland }\\
{$^d$ \small Physics Department, Technion, Haifa 32000, Israel}\\
\end{center}
\vspace*{\fill}
 
\begin{abstract}
We suggest to use operator product expansion and  
QCD sum rule techniques
to estimate the long-distance contribution
to the amplitude of the exclusive rare decay 
$B\rightarrow K^* \gamma$. In contrast to a
phenomenological description in terms of $\psi$ resonances
converting into a photon,
the virtual charm quark loop interacting  
with soft gluons is represented by a new
quark-gluon-photon operator. The matrix element 
of this operator is calculated in the same approximation 
as the matrix element of
the leading magnetic penguin operator.
The overall correction is found to be small, not more than
about $5\%$.

\end{abstract}
 
\vspace*{\fill}
 
\begin{flushleft}
\noindent$^1$ {\small on leave of absence from 
the Yerevan Physics Institute, 375036 Yerevan, Armenia}\\

\baselineskip=16pt                                 
\end{flushleft}

\section{Introduction}

Rare decays of $B$ mesons such as $B\rightarrow K^* \gamma$  
discovered at CLEO \cite{CLEO},
proceed mainly through loops, and are therefore an important tool 
to study new
physics, once the standard model contributions are 
sufficiently understood.
The generally accepted calculational 
procedure is to use an 
effective Hamiltonian 
\be
\label{heff}
{\cal H }_{eff} = 4\frac{G_F}{\sqrt{2}}V_{cb}V_{cs}^*
\sum_{i=1}^8C_i(\mu)O_i(\mu)
\ee
with suitable operators $O_i$ and Wilson coefficients $C_i$,
renormalized at the scale $\mu$. 
The task is then to evaluate the Wilson coefficients
and the matrix elements of the operators between the initial 
and final hadronic states, in the case at hand, between $B$ and $K^*$. 
The coefficients $C_i$ can be calculated perturbatively. A comprehensive
review of the current status of the Hamiltonian (\ref{heff})
can be found in \cite{BBL}.  
In order to obtain the 
exclusive matrix elements $\langle K^* \mid O_i \mid B \rangle$, 
some hadronic model or nonperturbative
method must be invoked. 

In leading 
logarithmic approximation, only  
the matrix element of the dominant $O_7$ operator is needed.  
It has been calculated using various methods
\cite{models,sumrules,Ball,ABS,lattice}.
However,  there remains a
rather strong $\mu$ dependence. It can be reduced by
next-to-leading order calculations of the $C_i$ and
a corresponding higher order in $\alpha_s$
determination of the matrix elements. 
For example, in the case of the 
inclusive reaction $b \rightarrow s \gamma$ 
one has to calculate the two-loop-diagrams 
involving the operator $O_2$. Here, 
the next-to-leading analysis has essentially been completed
\cite{AG,Pott,GHW,Misiak}. Generally, it is assumed that 
a similar approach is adequate for exclusive transitions. 
The corresponding diagrams are shown in Fig. 1.

When considering the penguin-like loops in these diagrams,
a perturbative calculation seems adequate for 
gluon virtualities $|k^2| \geq m_b^2$ 
and even when $m_c^2 \leq |k^2| \leq m_b^2 $. Furthermore,
if the photon is emitted from the $b$ or $s$ quark or from
the spectator quark (Figs. 1b,d), the integration over 
the $c$-quark loop produces  
a factor $k^2$ which cancels the denominator 
of the gluon propagator, effectively suppressing low-virtuality gluons.  
However, when the $c$--quark loop 
converts into a photon (Fig. 1a,c), 
the gluon may have a 
characteristic virtuality $|k^2|  \ll 4m_c^2$. This case is 
beyond perturbation theory.
The role of these long-distance penguins in  
$B\rightarrow K^*\gamma$  illustrated graphically in Fig. 2 
will be the main concern of this paper.

Their effect has been
considered by many authors on a phenomenological
level. The possibility extensively discussed in the literature 
\cite{psi} is to model the transition $B\rightarrow M \gamma$, $M$ being 
a light meson,  by a sum over 
the nonleptonic decays $B \rightarrow V^* M$  
followed by the transition $V^{*}\rightarrow \gamma$
where $V^{*}$ is an appropriate virtual
vector state such as the $J/\psi$.
Since at zero momentum squared $p^2$ of the photon, 
the characteristic distance to the physical threshold 
$p^2=M^2_\psi$ is rather large,
not only the ground state 
$J/\psi$ but also excited $\psi$ states must be
included. Unfortunately, the transverse parts of these 
amplitudes are poorly known
and the extrapolation from $p^2=M^2_\psi$ 
to $p^2=0$ is not straightforward. Therefore, for a real photon, 
the vector-resonance description 
of the virtual $\bar{c}c$ loop seems uncertain, 
calling for a more reliable or at least 
different estimate.  

As a new approach to the problem, 
we suggest to use operator product expansion (OPE) and QCD sum rules
\cite{SVZ}. Several important circumstances make these methods 
exceptionally suitable. 
Firstly,  the fact that the $\bar{c}c$ loop is far off--shell,   
which is a disadvantage for $\psi$ insertion,
turns out to be a major advantage here. It 
means that we must keep only a few terms in the 
OPE. Secondly, the QCD sum rule method 
provides a physically very intuitive
parametrization of the interaction of the $\bar{c}c$ loop 
with soft gluons in terms of vacuum condensates.
Thirdly, and most importantly, we will be able to compare 
the long-distance effect in $B\rightarrow K^* \gamma$ with the 
contribution of the dominant magnetic penguin operator 
obtained in one and the same calculational framework.
Of course, as with any low energy effective method, there are some
unavoidable theoretical uncertainties. One 
of them is the question of double counting in the perturbative 
and nonperturbative contributions. As usual,
we assume that the former are small in the region where the
latter dominate, that is at $|k^2|\ll 4m_c^2$. 

There are other nonperturbative effects, such as 
those illustrated in Fig. 3, which involve radiative matrix  elements of
the four-fermion operators $O_{1,2}$. These contribute
already at tree level and have been  
estimated \cite{KSW,AB} using QCD sum rules on the light-cone. 
They are unimportant for the $b \rightarrow s$
transitions due to CKM suppression (Fig. 3a), and also  relatively small 
for $b \rightarrow d$ transitions (Fig. 3b).

The paper is organized as follows. In Sect. 2, we consider
the contribution of the operator $O_2$ 
to the exclusive $B\rightarrow K^*\gamma$ amplitude 
and single out the term responsible for soft gluon
emission from the $c$-quark loop. 
With the help of OPE, this term is expressed by a composite 
operator involving quark, gluon and photon fields. In Sect. 3, 
we derive a QCD sum rule for the $B \rightarrow K^* \gamma $ 
matrix element, including the new operator. 
The most important nonperturbative contributions 
of gluon, quark-gluon, and quark condensates up to dimension 6  
to the relevant correlation function are calculated 
in Sect. 4. Our numerical results are presented in Sect. 5 and our
conclusions in Sect. 6.

\section{Long-distance contribution of the operator $O_2$ }

In the framework of the effective 
Hamiltonian (\ref{heff}) the exclusive $B \rightarrow K^* \gamma $
decay amplitude  is given by 
$$
A(B \rightarrow K^* \gamma)= \langle K^*(q)~ \gamma(p) \mid 
{\cal H}_{eff}\mid  B(p+q) \rangle =
$$
\be 
= \frac{4G_F}{\sqrt{2}}V_{cb}V_{cs}^*\langle K^*(q)\mid [ C_7O_7
+i\epsilon^\mu \sum_{i\neq 7}C_i
\int d^4x~e^{ipx} T \{j_\mu^{em}(x)O_i(0)
\}]\mid B(p+q) \rangle ~,
\label{ampl}  
\ee
where the four-momenta of the $B$ and 
$K^*$ meson, and the photon are denoted by 
$p+q$, $q$, and $p$, respectively,
while $\epsilon_\mu$ is the photon polarization 
four-vector. The electromagnetic current is given by
\be
j_\mu^{em}= \sum_{q=u,d,s,c,b}e_q \bar{q}\gamma_\mu q~~,
\label{curr}
\ee
where $e_q$ are the quark electric charges. 

In (\ref{ampl}), we have isolated the operator $O_7$, as it is the only
operator which contains the photon field at the tree level:
\be
O_7= \frac{em_b}{16 \pi ^2}\bar s_L\sigma_{\mu \nu}
b_R F^{\mu\nu}~,
\label{o7}
\ee
with $q_{R,L}= \frac12(1\pm \gamma_5)q$, $F_{\mu\nu}$ 
being the photon field tensor, 
and $e=\sqrt{4\pi\alpha}$
being the electromagnetic coupling. 
Note that throughout the paper we put $m_s=0$.
The other operators generate radiative decays when combined with the 
electromagnetic interaction of quarks with the photon.
In this paper, we restrict ourselves to the four-quark 
operators 
\be
O_1= (\bar c_L\gamma_\mu c_L)(\bar s_L\gamma^\mu b_L)~,
\label{o2}
\ee
\be
O_2= (\bar s_L\gamma_\mu c_L)(\bar c_L\gamma^\mu b_L)~,
\label{o1}
\ee
relying on the fact that the other 
operators have much smaller short-distance coefficients. 
To our knowledge, an analysis
of long-distance effects in $B\rightarrow K^* \gamma$ 
generated by $O_{3-6}$ 
and $O_8$ has never been done and is certainly desirable.
Perturbative contributions of $O_8$ have been estimated
in the relativistic quark model \cite{CM} and found to be small.  

For $O_{2}$, the 
relevant diagrams contributing to the amplitude (\ref{ampl})
are  shown in Fig. 1. 
The analogous graphs for the operator $O_1$ vanish 
for a real photon. 
After Fierz transformation,
the operator $O_2$ reads: 
\bq
O_2= \frac{1}3 O_1 + \frac12\tilde{O}_1~,
\label{o11}
\\
\tilde{O_1}= 
(\bar c\Gamma_\mu \frac{\lambda^a}2 c)
(\bar s\Gamma^\mu \frac{\lambda^a}2 b)~,
\label{otilde}
\eq
where $\Gamma_\mu= \gamma_\mu(1-\gamma_5)$.  
It is only the operator $\tilde{O}_1$ which plays a role. 
The effective colour transfer in the weak vertex 
is compensated by a gluon 
exchange between the $c$--quark loop and 
other quarks in the process. 
The photon emission from the $c$--quark loop 
is therefore a purely nonfactorizable effect.

Keeping in (\ref{ampl}) only the operators $O_7$ and  $\tilde{O}_1$  
and retaining in $j^{em}_\mu$ 
only the $c$--quark part, 
the decay amplitude simplifies to 
$$
A(B \rightarrow K^* \gamma)
= \frac{4G_F}{\sqrt{2}}V_{cb}V_{cs}^*\langle K^*(q)\mid \Big[ C_7O_7
$$
\be 
+i\frac{e_c C_2}2\epsilon^\mu \int 
d^4x ~e^{ipx} T\{\bar{c}(x)\gamma_\mu c(x)\tilde{O}_1(0)\}
\Big]\mid B(p+q) \rangle .
\label{ampl1}  
\ee
In this expression, one encounters
a product of $\bar{c}\gamma_\mu c$ and 
$\bar c\Gamma_\rho \frac{\lambda^a}2 c$
currents generating a $c$-quark loop. 
When the photon is on-shell, the 
$c$-quark pair is far off-shell. 
In this situation, we expect that the OPE for the $c$--quark loop 
in terms of local operators is adequate. 

The first term of this expansion
is simply a gluon field operator.
The corresponding Wilson coefficient
is represented by the diagrams of Fig. 4. It can easily be 
calculated using the fixed-point ($x=0$) gauge 
which in the approximation needed reads
\be
A^a_{\mu}(x)=1/2x^{\tau}G^a_{\tau\mu}(0) ~.  
\label{gauge}
\ee
The OPE then results in   
\be
\int e^{ipx}T\{\bar{c}(x)\gamma_\mu c(x), \bar{c}(0)
\Gamma_\rho\frac{\lambda^a}2 c(0)\}
= C_{\alpha\beta\mu\rho}g\tilde{G}^{a\alpha\beta}+ ...
\label{charm}
\ee
with the short-distance coefficient ($p^2 = 0$), 
\be 
C_{\alpha\beta\mu\rho}=
-\frac{i}{48\pi^2m_c^2}\left(
p_\alpha p_\rho g_{\beta\mu} +p_\alpha p_\mu g_{\beta\rho}\right)~,
\label{C}
\ee
and $\tilde{G}^{a\alpha\beta}= 
1/2\epsilon^{\alpha\beta\tau\lambda}G_{\tau\lambda}^a$.
Note that in the diagrams of Fig. 4 only the axial part
of $\Gamma_\rho\lambda^a$ contributes, whereas the 
contribution of the vector part vanishes due to 
Furry's theorem.
The higher dimension operators in the expansion (\ref{charm}) 
denoted by the ellipses contain derivatives of the gluon field
and/or additional gluon field operators. They are neglected here.
Substituting the expansion (\ref{charm}) into  
(\ref{ampl1}), and putting $e_c= (2/3)e$ yields
\be
A(B \rightarrow K^* \gamma)=
4\left(\frac{G_F}{\sqrt{2}}\right)V_{cb}V_{cs}^*
\langle K^*(q)\mid [ 
C_7 O_7 +C_2 O_F] 
\mid B(p+q) \rangle 
\label{ampl2}
\ee
where 
\be
O_F=-\frac{e}{288\pi^2m_c^2}\bar{s}\Gamma_\rho\frac{\lambda^a}2g\tilde
{G}_{\alpha\beta}^a D^\rho F^{\alpha\beta}b  
\label{eff}
\ee
is a new effective operator describing 
the soft gluon interaction sketched in Fig. 2.
 
Before turning to the estimate of the matrix element of $O_F$,
a comment on the accuracy of approximation (\ref{charm}) is in order.
A similar approximation was used in \cite{BS}
in estimating the nonfactorizable contributions
to $B \rightarrow D \pi$  
(see also \cite{Halperin}). In this analysis, the criteria 
for the applicability of OPE are discussed in detail.
Following the same line of arguments, we notice that the
next-to-leading operator in (\ref{charm}) will contain
a derivative of the gluon field $D_\alpha\tilde{G}$
multiplied by the photon momentum $p_\alpha$ and an extra   
factor of order of $1/4m_c^2$. The overall suppression
factor  will thus be of order of $\mu_{h} p_0/4m_c^2$ 
where $\mu_{h}$ is a characteristic hadronic scale, 
and $p_0 \simeq m_b/2$ is the photon energy. Assuming 
$\mu_{h} \simeq 300 $ MeV, $m_b\simeq$ 5 GeV
and $m_c \simeq 1.3$ GeV one expects a 10\% effect. From this 
very rough estimate the truncation of (\ref{charm})
after the first term appears justifiable. 
On the other hand, one clearly cannot apply 
the same approximation to light-quark loops. 

Next, we write the matrix element of
the magnetic penguin operator in the form 
\be
\langle K^*(q)\mid  O_7
\mid B(p+q) \rangle 
=
\frac{em_b}{32\pi^2}
F\left(i\epsilon_{\alpha \beta \sigma \tau} q^\tau
-2g_{\sigma\alpha} q_\beta\right)e_{K^*}^\sigma F^{\alpha\beta} .
\label{penguin}
\ee
where $e_{K^*}$ is the polarization vector of the $K^*$
and $F_{\mu\nu}= i(e_\nu p_\mu-e_\mu p_\nu)$. 
In the above,  
the relation
\be
\sigma_{\mu\nu}\gamma_5 = -\frac{i}2\epsilon_{\mu\nu\alpha\beta}
\sigma^{\alpha\beta} 
\label{sigma}
\ee
has been used, 
which leaves only one independent invariant amplitude $F$.
Similarly to (\ref{penguin}), we parametrize the  
matrix element (\ref{ampl2}) of the operator $O_F$ 
in the form 
\be
\langle K^*(q)\mid O_F\mid B(p+q)\rangle 
=-\frac{e}{288\pi^2m_c^2}
\left(iL\epsilon_{\alpha \beta \sigma \tau} q^\tau
-2\tilde{L}g_{\sigma\alpha} q_\beta\right)e_{K^*}^\sigma F^{\alpha\beta} ~.
\label{matrix}
\ee
Note that the invariant amplitudes $L$ and $\tilde{L}$ have dimension GeV$^3$
and that, in general, $\tilde{L}\neq L$. 
Finally, substituting (\ref{matrix}) and (\ref{penguin}) 
in (\ref{ampl2}) one gets
$$
A(B \rightarrow K^* \gamma)
= \left(\frac{eG_FV_{cb}V_{cs}^*}{8\sqrt{2}\pi^2}\right)
\Bigg[-i\epsilon_{\alpha \beta \sigma \tau}q^\tau
\left(-C_7m_bF+\frac{C_2}{9m_c^2}L\right)
$$
\be
+2g_{\sigma\alpha} q_\beta
\left(-C_7m_bF+\frac{C_2}{9m_c^2}\tilde{L}\right)
\Bigg]e_{K^*}^\sigma F^{\alpha\beta} .
\label{amplitude}
\ee
While the amplitude proportional to $F$ has been 
investigated previously, the long-distance corrections proportional
to $L$ and $\tilde L$ are new.

\section{ QCD sum rule for the $B\rightarrow K^*\gamma$
amplitude}   
Estimates for the  matrix element (\ref{penguin}) already exist 
in the framework of local 
\cite{sumrules,Ball} and light-cone QCD sum rules \cite{ABS}. 
The additional matrix element 
(\ref{matrix})
can in principle be calculated in the same framework with the 
same accuracy.
The light-cone approach to this  
matrix element involves quark-antiquark-gluon 
wave functions of the $K^*$ meson.
Since these wave functions are not yet sufficiently 
known, we shall estimate (\ref{matrix}) using the local OPE. 
Following \cite{sumrules,Ball}, we consider 
the three-point correlation function 
$$
T_{\alpha \beta}^{\nu}(p,q)F^{\alpha\beta} = 
- \frac{e}{32\pi^2}\int d^4xd^4y\exp{[-i(p+q)x+iqy]} 
\langle 0|T\Big\{\bar{u}(y)\gamma^{\nu} s(y),
$$
\be
[C_7m_b\bar{s}(0)\sigma_{\alpha \beta}(1+\gamma_5)b(0)
-\frac{iC_2}{9m_c^2} p_\rho\bar{s}(0)\Gamma^\rho\frac{\lambda^a}
2g\tilde{G}_{\alpha\beta}^a (0)b(0)] 
,\bar{b}(x)i\gamma_5 u(x)\Big\}|0 \rangle F^{\alpha\beta} ,
\label{corr}
\ee
interpolating the complete matrix element in (\ref{ampl2}), 
with the $B$ and $K^*$ mesons replaced by appropriate generating 
currents. Moreover, for the operators $O_7$ and $O_F$ 
we have substituted the explicit expressions 
(\ref{o7}) and (\ref{eff}). 
Isolating in (\ref{corr}) 
the tensor structures (\ref{penguin}) and (\ref{matrix}) 
we write 
\begin{equation}
T_{\alpha \beta}^{\nu} F^{\alpha \beta}=
\frac{e}{32\pi^2}
\left(-Ti \epsilon_{\alpha\beta\nu\tau} q^{\tau} F^{\alpha\beta}
+2\tilde{T} q_{\beta} F_{\nu\beta}+ ...\right) ~,
\label{tensor}
\end{equation}
where the invariant amplitudes $T$ and $\tilde{T}$ are functions 
of the two independent variables $(p+q)^2$ and $q^2$. 

Inserting 
in the r.h.s. of (\ref{corr}) the complete sets of hadronic states carrying  
the $B$ and $K^*$  quantum numbers one obtains the double dispersion 
relation
\bq
T((p+q)^2,q^2)= \frac{m_{B}^2 m_{K^*}f_{B}f_{K^*}}{ m_b(m_B^2-(p+q)^2)
(m_{K^*}^2-q^2)}
\left(C_7m_bF-\frac{C_2}{9m_c^2}L\right)
\nonumber
\\
+ \int d s_1ds_2 \frac{\rho_h(s_1,s_2)}{
(s_1-(p+q)^2)(s_2-q^2)}  
\label{t1}
\eq
where we have used (\ref{penguin}) and (\ref{matrix}) together 
with
\be
\langle B \mid \bar{b}i\gamma_5 u \mid 0 \rangle= \frac{m_B^2}{m_b}f_B ~,
\label{fB}
\ee
\be
\langle 0 \mid \bar{u}\gamma_\nu s \mid K^{*}\rangle= 
m_{K^*}f_{K^*}e_{K^*\nu}  ~. 
\label{fK}
\ee
The integral 
over the double spectral density $\rho_h$ in (\ref{t1}) represents 
the contributions of excited $B$ and $K^*$ states as well as of continuum
contributions. Possible subtraction terms are omitted for brevity.   
The analogous dispersion relation for $\tilde{T}$ is
obtained from (\ref{t1}) by the replacements
$L\rightarrow \tilde{L}$ and
$\rho_h\rightarrow \tilde{\rho}_h$ .

Furthermore, at $p^2=0$ and at large spacelike 
values of $(p+q)^2$ and $q^2$, 
the amplitudes
$T$ and $\tilde{T}$ can be evaluated in QCD in terms of 
vacuum expectation values of certain local operators $ \Omega_d$: 
\be
T_{QCD}((p+q)^2,q^2) = 
\sum _{d} 
\left 
[C_7m_bT_{d}((p+q)^2,q^2)-\frac{C_2}{9m_c^2}U_{d}((p+q)^2,q^2)\right ]
\langle {\Omega}_d \rangle ~.
\label{expan}
\ee
A similar representation holds for $\tilde{T}_{QCD}$ with $U_d$ 
replaced by $\tilde{U}_d$. 
The index $d$ refers to the dimension of the operators.
In perturbation theory, only the $d=0$ unit operator 
contributes to (\ref{expan}). The nonperturbative effects are accumulated
in terms with $d>0$. In this analysis, we take into account 
$\langle {\Omega}_3 \rangle=\langle \bar{q}q \rangle $,
$\langle {\Omega}_4 \rangle = 
\langle \frac{\alpha_s}{\pi}G^a_{\mu\nu}G^{a\mu\nu} \rangle$, 
$\langle {\Omega}_5 \rangle = m_0^2\langle\bar{q}q \rangle$,
and $\langle {\Omega}_6 \rangle =\langle \bar{q}q \rangle^2$.
These are the standard parametrizations for gluon and 
quark-gluon condensate densities. 
For the four-quark condensate the vacuum insertion
approximation \cite{SVZ} is used. 

Following the usual procedure, one equates (\ref{t1})
and (\ref{expan}), and replaces the double spectral density 
$\rho_h$ in (\ref{t1}) by the double imaginary part of 
the amplitude $T_{QCD}$. Finally, the  
Borel transformation
\begin{equation}
\hat{B}_{M^2} f(Q^2)\equiv \lim_{Q^2\to\infty,n\to\infty,Q^2/n=M^2}
\frac{(Q^2)^{n+1}}{n!}
\left(-\frac{d}{dQ^2}\right)^n f(Q^2) \equiv f(M^2) ~ 
\label{borel}
\end{equation}  
with respect to the variables $(p+q)^2$ and $q^2$ is applied,
in order to suppress exponentially the  
contribution of higher states and to get rid of subtraction terms
in the dispersion relations.
The second invariant amplitude $\tilde{T}$ is treated in the same way.

This yields the sum rule 
$$
C_7m_bF-\frac{C_2}{9m_c^2}L
= \frac{m_b}{m_{B}^2m_{K^*}f_{B}f_{K^*}}
e^{m_B^2/M_1^2 + m_{K^*}^2/M_2^2}
$$
$$
\times \Bigg \{\frac{1}{\pi^2}\int\int^{\{s_0^B, s_0^{K^*}\}} 
ds_1 ds_2 \left[C_7m_bIm_{s_1}Im_{s_2}T_{0}(s_1,s_2)
-\frac{C_2}{9m_c^2}Im_{s_1}Im_{s_2}U_{0}(s_1,s_2)\right]
e^{-s_1/M_1^2 -s_2/M_2^2}
$$
\be
+\sum_{d\neq 0}\left[C_7m_bT_{d}(M_1^2,M_2^2)
-\frac{C_2}{9m_c^2}U_{d}(M_1^2,M_2^2) \right]
\langle {\Omega}_d \rangle 
\Bigg\}~.   
\label{sr}
\ee  
where the threshold parameters
$s_0^B$ and $s_0^{K^*}$ indicate the subtraction 
of higher state contributions.
A second sum rule is obtained
from (\ref{sr}) by replacing $L \rightarrow \tilde{L}$ 
and $U_d \rightarrow \tilde{U}_d$.

In the approximation considered, the above sum rule can obviously
be divided into two separate 
sum rules, one for $F$ (terms involving $T_d$ ) 
and one for $L$ (terms involving $U_d$). 
The sum rule for $F$ have already been analyzed 
\cite{Ball} including terms up to $d=6$, in 
lowest order in $\alpha_s$.
In the following, we calculate the sum rule for the long-distance 
corrections $L$ and $\tilde{L}$ in the {\it same} approximation.

\section{Calculation of the long-distance amplitudes} 

We need to calculate the coefficients $U_d$ and $\tilde{U}_d$  
up to $d=6$. The coefficients 
$U_0$, $\tilde U_0$, $U_3$ and $\tilde U_3$
are of order $\alpha_s$ as is obvious from 
Fig. 5. Since 
the $O(\alpha_s)$  corrections to the sum rule for $F$
(exemplified in Fig. 5a,d) are not taken into account
in \cite{Ball} we put $U_0=\tilde U_0=U_3=\tilde U_3=0$
in (\ref{sr}).
A systematic inclusion of  these effects would  
require two-loop calculations and 
a proper treatment of operator mixing. 
Thus, to the requested accuracy, 
we have 
$$
L = \frac{m_b}{m_{B}^2m_{K^*}f_{B}f_{K^*}}
e^{m_B^2/M_1^2 + m_{K^*}^2/M_2^2}
\Big\{
U_4(M_1^2,M^2_2)\langle \frac{\alpha_s}{\pi}G^a_{\mu\nu}G^{a\mu\nu} \rangle 
$$
\be
+
U_5(M_1^2,M^2_2)m_0^2\langle \bar{q}q \rangle 
+
U_6(M_1^2,M^2_2)\langle \bar{q}q \rangle^2 
\Big\}~,
\label{sr1}
\ee
and a similar expression for the 
amplitude $\tilde{L}$ with $U_{4,5,6}$  replaced by 
$\tilde{U}_{4,5,6}$.

In order to calculate the  
Wilson coefficients $U_{4}$  
and  $\tilde{U}_{4}$, 
one has to contract all quark fields in the second term
of the correlation function (\ref{corr}).
One of the three resulting quark propagators is substituted 
by the first order expression 
in the external gluon field $A^a_{\mu}$, while for the other two
propagators the free approximation is used.
This yields the diagrams of Fig. 6a, b, c.
The calculation of these diagrams is performed 
in the Fock-Schwinger gauge (\ref{gauge}).
Furthermore, when expressing the result in form of a double
dispersion integral in the variables
$(p+q)^2$ and $q^2$ one may omit all terms which depend only on one
of these variables since such terms vanish after 
Borel transformation. The 
double dispersion contribution of  
the diagram Fig. 6c vanishes altogether. The result is  
\be
U_4=U_4^{(Fig.6a)}+U_4^{(Fig.6b)}~,
\label{u4}
\ee
with 
$$
U_4^{(Fig.6a)}=-\frac{m_b}{48}
\int_{m_b^2}^\infty \frac{ds_1}{s_1-(p+q)^2}
\int_0^{s-m_b^2}\frac{ds_2}{s_2-q^2}\left[1-\frac{m_b^2}{s_1-s_2}\right]
\left[\frac{1}{s_2-q^2}+\frac{1}{s_1-s_2}\right]~,
$$
\be
U_4^{(Fig.6b)}=\frac{m_b^3}{48}
\int_{m_b^2}^\infty \frac{ds_1}{s_1-(p+q)^2}
\int_0^{s-m_b^2}\frac{ds_2}{(s_2-q^2)(s_1-s_2)}
\left[\frac{1}{s_1-(p+q)^2}+\frac{1}{s_2-q^2}\right]~,
\label{UU}
\ee
and 
\be
\tilde{U}_4=\tilde{U}_4^{(Fig.6a)}+\tilde{U}_4^{(Fig.6b)}~,~~ 
\label{tu4}
\ee
with
\be
\tilde{U}_4^{(Fig.6a)}=-U_4^{(Fig.6a)}~,~~
\tilde{U}_4^{(Fig.6b)}= U_4^{(Fig.6b)}~.
\label{uu4}
\ee

The coefficients $U_5$ and $\tilde{U}_5$ 
of the $d=5$ quark-gluon condensate term are obtained from 
Fig. 6d which involves the vacuum average 
of the $\bar{u}uG $ operator. The analogous diagram,
involving  $\bar{s}sG$ depends only on $q^2$ and can therefore
be dropped. One finds
\be
U_5=\tilde{U}_5=
-\frac{1}{12}
\frac{m_b^2}{(m_b^2-(p+q)^2)(-q^2)}~. 
\label{u5}
\ee

There are two  $d=6$ contributions to the correlation
function (\ref{corr}). One of them originates  from 
expanding $\bar{u}(y)$ and $u(x)$ at $x=y=0$. 
In vacuum insertion approximation, 
the vacuum average of the resulting operator 
$\bar u \nabla_\rho u G^a_{\mu\nu} $ is expressed 
in terms of $\langle \bar q q \rangle^2$ according to the 
formula given in \cite{IoffeSmilga}. The 
corresponding coefficients derived from Fig. 6d read 
\be
U_6^{(Fig.6d)} = \tilde{U}_6^{(Fig. 6d)}
=\frac{4\pi\alpha_sm_b}{27}
\frac{1-m_b^2/q^2}{(-q^2)
(m_b^2 -(p+q)^2)}~.
\label{u6}
\ee
The second $d=6$ contribution 
comes from the four-quark condensate. The relevant 
diagrams are shown in Fig. 6e and 6f. 
The coefficients are
$$
U_6^{(Fig.6e,f)}=0~, 
$$
\be
\tilde{U}_6^{(Fig.6e,f)} 
= \frac{4\pi\alpha_s m_b}{9}
\frac{1-m_b^2/q^2}{(-q^2)
(m_b^2 -(p+q)^2)}~.
\label{u66}
\ee
In total, we have 
\be
U_6=U_6^{(Fig.6d)}~, 
\label{U6}
\ee
\be
\tilde{U}_6=\tilde{U}_6^{(Fig. 6d)}+\tilde{U}_6^{(Fig. 6e,f)}~.
\label{tU6}
\ee

After double Borel transformation (\ref{borel}) 
of the above expressions for $U_d$ and $\tilde{U}_d$ 
in $(p+q)^2$ and $q^2$,  
and substitution  in (\ref{sr1}) and  
in the analogous expression for $\tilde{L}$,   
we obtain the final sum rules

$$
L= \frac{m_b}{m_{K^*}m_B^2f_Bf_{K^*}}
\exp\left(\frac{m_B^2}{M_1^2}+\frac{m_{K^*}^2}{M_2^2}\right)
$$
$$
\times\Bigg \{ \frac{m_b}{48}\langle \frac{\alpha_s}{\pi}GG
\rangle \int_{\frac{m_b^2}{M_1^2}}^\infty d\tau\exp(-\tau)
\Big[\frac{m_b^2}{\tau}-\frac{M_1^4}{M_1^2 + M_2^2}
\left(1-\frac{m_b^2}{\tau M_1^2}\right)
\left(1+\frac{M_2^2}{\tau M_1^2}\right)\Big]
$$
\be
-\Big[\frac{m_0^2\langle\bar{q}q \rangle m_b^2}{12}-
\frac{4\pi \alpha_s\langle \bar{q} q \rangle^2 m_b}{27}
\left(1+\frac{m_b^2}{M_2^2}\right)
\Big]\exp\left(-\frac{m_b^2}{M_1^2}\right) 
\Bigg\}~,
\label{L}
\ee

$$
\tilde{L}= \frac{m_b}{m_{K^*}m_B^2f_Bf_{K^*}}
\exp\left(\frac{m_B^2}{M_1^2}+\frac{m_{K^*}^2}{M_2^2}\right)
$$
$$
\times\Bigg\{ \frac{m_b}{48}\langle \frac{\alpha_s}{\pi}GG
\rangle \int_{\frac{m_b^2}{M_1^2}}^\infty d\tau\exp(-\tau)
\Big[\frac{m_b^2}{\tau}+\frac{M_1^4}{M_1^2 + M_2^2}
\left(1-\frac{m_b^2}{\tau M_1^2}\right)
\left(1+\frac{M_2^2}{\tau M_1^2}\right)\Big]
$$
\be
-\Big[\frac{m_0^2\langle\bar{q}q \rangle m_b^2}{12}-
\frac{16\pi \alpha_s\langle \bar{q} q \rangle^2 m_b}{27}
\left(1+\frac{m_b^2}{M_2^2}\right)
\Big]\exp\left(-\frac{m_b^2}{M_1^2}\right) 
\Bigg\}~.
\label{LL}
\ee

\section{Numerical estimates}

For the numerical estimate  of the amplitudes $L$ and $\tilde{L} $
we take the values of the physical 
parameters also used in \cite{Ball}, that is  
$ m_b= 4.8 $ GeV, $\langle \frac{\alpha_s}{\pi} GG \rangle =0.012$ GeV$^4$,
$\langle \bar{q} q \rangle= -(240\mbox{MeV})^3$ 
at the scale $\mu=1$ GeV, $m_0^2=0.8$ GeV$^2$, $\Lambda_{QCD}^{(4)}= 230$ MeV,
$f_{K^*}= 210$ MeV, and $f_B=140$ MeV. The latter value results 
from the corresponding two-point sum rule without $O(\alpha_s)$ corrections
and should not be confused with the physical value of $f_B$ which is larger.   
We also use the same interval of the Borel mass $M_1$,  
$7 \leq M_1^2 \leq 10 $ GeV$^2$, keeping the ratio 
$M_1^2/M_2^2 = 3$ fixed.  
The characteristic scale at which the amplitudes  
are estimated by the sum rules is 
of order of the average Borel parameter 
that is, of order
of the virtuality of the quarks in (\ref{corr}). 
In \cite{Ball}, the scale $\mu_{B}=(M_1M_2)^{1/4} \simeq 2$ GeV
is chosen for evaluation of $F$. 
Neglecting an inessential evolution of the quark-gluon condensate 
and of the product $\alpha_s\langle \bar{q} q \rangle^2 $ 
from $\mu= 1 $ GeV to $\mu=\mu_{B}$, the sum rules 
(\ref{L}) and (\ref{LL}) yield: 
$$
L = 0.55 \pm 0.1 \mbox{ GeV}^3~,
$$
\be
\tilde{L}= 0.70 \pm 0.1 \mbox{ GeV}^3 ~.
\label{l1}
\ee

The theoretical uncertainty quoted here reflects only the variation
of $L$ and $\tilde{L}$ with the 
Borel masses $M_1$ and $M_2$ in the accepted interval. 
In contrast to \cite{Ball}, we have neglected 
the finite 
strange quark mass effect which however makes no difference numerically.   
The dominant contributions in (\ref{L}) and (\ref{LL}) come from the
quark-gluon condensate (75 \%) and gluon condensate
(15-20 \%), while  the  $d=6$ terms do not
exceed 10\%.   
The deviation of $\tilde{L}$ from  $L$ which is numerically 
not very significant is generated partly by the gluon condensate 
and partly by the four-quark condensate. 

For comparison, we quote the corresponding prediction 
for the penguin amplitude (at $\mu=\mu_B$) given in \cite{Ball}: 
\be
F= 0.9 \pm 0.1~. 
\label{f}
\ee
With (\ref{l1}) and (\ref{f})
we find for the relative magnitude of the 
long-distance corrections 
in (\ref{amplitude}): 
$$
r= \left(\frac{C_2}{-9C_7}\right)
\frac{L}{m_b m_c^2 F} \simeq 0.02 \div 0.04 ~,
$$
\be
\tilde r= \left(\frac{C_2}{-9C_7}\right)
\frac{\tilde L}{m_b m_c^2 F} \simeq 0.03 \div 0.05 ~,
\label{correction}
\ee
Apart from $m_b=4.8$ GeV and $m_c= 1.3 ~\mbox{GeV}$ we have used here
$C_2 = 1.14 $ and $C_7=-0.30$,
that is the leading-logarithmic approximation
of the short-distance coefficients   
at $\mu\simeq m_b$ \cite{BBL}.
Since both estimates (\ref{l1}) and (\ref{f})
are obtained by the same method and in the same approximation, 
one can hope that many of the uncertainties cancel in the above ratios.

\section{Conclusion}

In this paper, we have estimated the long-distance effect 
in $B\rightarrow K^* \gamma$
due to interactions of the virtual $c$-quark loop with soft gluons.
Using OPE we have derived a new quark-gluon-photon operator $O_F$ given 
in (\ref{eff}) which is responsible for the effect. 
Furthermore, we have obtained a QCD sum rule for the matrix 
element  $\langle K^* \gamma \mid O_F\mid B\rangle $
taking into account  
the most important nonperturbative contributions. 
Quantitatively, the correction is found to increase the   
magnetic penguin amplitude by a few percent.
This result indicates that long-distance 
effects do not play an important role in the 
exclusive decay $B\rightarrow K^* \gamma$.

The estimate presented here is not free  
from theoretical uncertainties.
They are partly due to the lack of knowledge of the
perturbative $O(\alpha_s)$ corrections 
to the Wilson coefficients. Related to that is the ambiguity in 
the choice of the scale $\mu$. In addition,
there are uncertainties inherent to the short-distance sum rules. 
It would therefore be desirable to 
repeat the calculation using  
light-cone sum rules for a cross-check. However, the latter approach 
requires a better understanding 
of the higher twist light-cone wave functions of the $K^*$ meson.

When this work was completed, the 
paper \cite{Voloshin} appeared in which the nonperturbative 
correction to the  {\it inclusive} $B \rightarrow X_s \gamma$
rate due to soft gluon emission by the $c$-quark loop 
is estimated. The responsible effective operator 
obtained in \cite{Voloshin} coincides with our operator 
(\ref{eff}). After revision of \cite{Voloshin},
the effect predicted for the inclusive rate  is similar in size to 
the correction found here for the 
exclusive channel $ B \rightarrow K^* \gamma $, however, 
opposite in sign.

\section{Acknowledgements} 
One of the authors ( A.K.) 
is grateful to P. Ball for a useful discussion 
concerning her work \cite{Ball}.
The work of A.K. and R.R. is supported by the Bundesministerium f\"ur Bildung 
und Forschung (BMBF) under contract 05 7WZ91P(0) and by the EC 
program HCM under contract CHRX-CT93-0132.
The work of G.S. and D.W is supported by Schweizerischer Nationalfond. 
\pagebreak

\newpage
\begin{figure}[htb]
\epsfysize = 8 cm
\epsffile{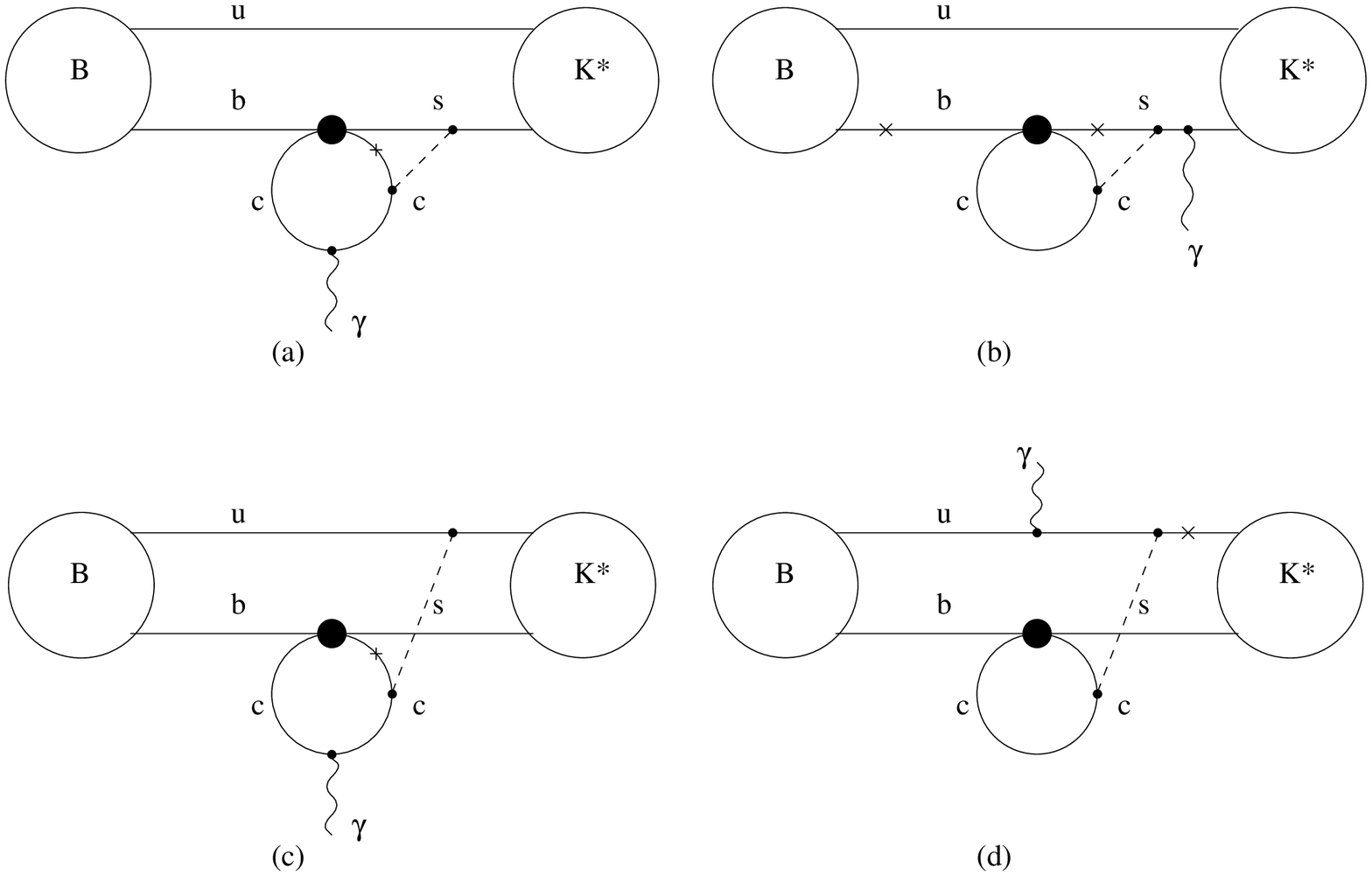}
\caption{Diagrams contributing to  
the $B\rightarrow K^*\gamma$ amplitude generated
by the operator $O_2$ (black circle).}
\end{figure}

\begin{figure}[htb]
\epsfysize = 3 cm
\epsffile{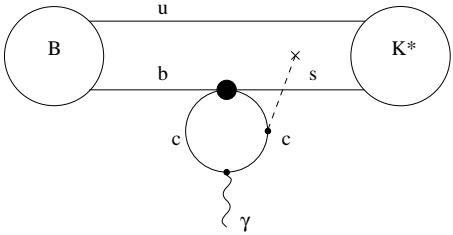}
\caption{Interaction of the virtual $c$-quark 
loop with soft gluons.} 
\end{figure}

\begin{figure}[htb]
\epsfysize = 2.3 cm
\epsffile{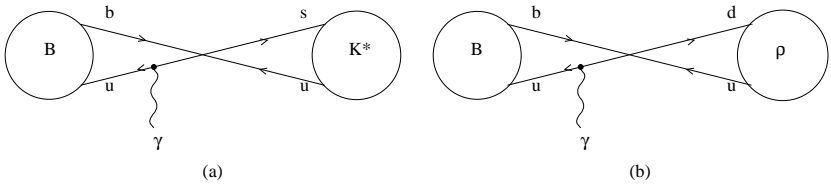}
\caption{Photon emission by quarks 
at long-distances
in (a) $B\rightarrow K^*\gamma$ and (b) $B\rightarrow \rho \gamma$.}
\end{figure}

\begin{figure}[htb]
\epsfysize = 2.5 cm
\epsffile{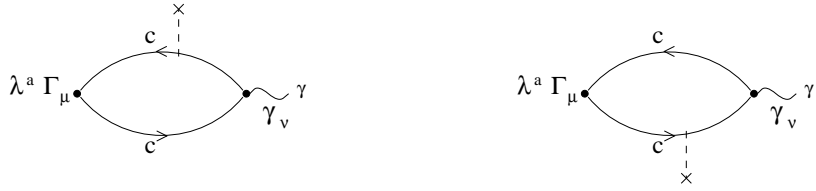}
\caption{Diagrams contributing to the operator product expansion
of the $c$-quark loop.}
\end{figure}

\begin{figure}[htb]
\epsfysize = 6 cm
\epsffile{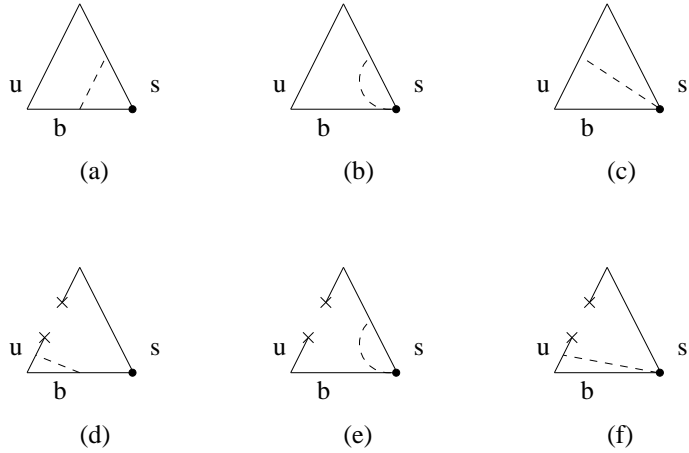}
\caption{Diagrams determining the $O(\alpha_s)$ corrections 
to the Wilson coefficients of the three-point correlation function
(\ref{corr}). 
The $\bar s b $ vertex involves $O_7$,
while the $\bar s b G$  vertex involves $O_F$.}
\end{figure}

\begin{figure}[htb]
\epsfysize = 6 cm
\epsffile{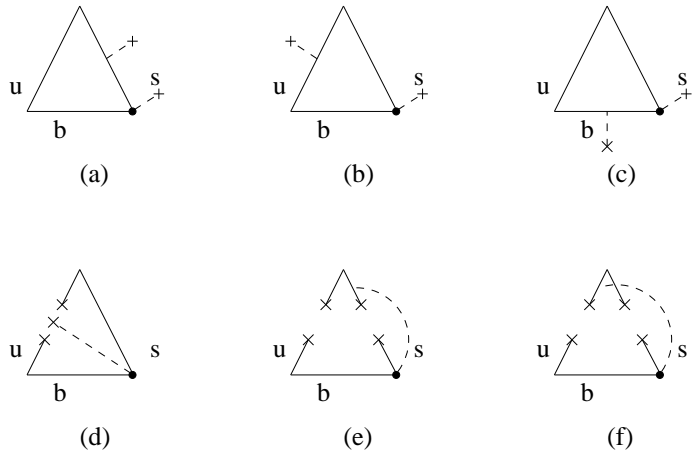}
\caption{
Diagrams yielding the coefficients 
(a-c) $U_4$, $\tilde{U_4}$ , (d) $U_5$, $\tilde{U_5}$,  
and (d-f) $U_6$, $\tilde{U_6}$ 
in the sum rules for $L$ and $\tilde{L}$.}
 
\end{figure}

\end{document}